\documentclass[12pt,a4paper]{article}
\pdfoutput=1
\usepackage{graphicx}
\usepackage{xcolor}
\usepackage{fullpage}
\newcommand{\ilde}{\tilde}
\newcommand{\up}{\uparrow}
\begin{document}
\title{The N(1440) Roper resonance in the nuclear model with explicit
mesons}
\author{
D.V.~Fedorov\footnote{Aarhus University, fedorov@phys.au.dk}
}
\date{}
\maketitle

	\begin{abstract}
We show that the N(1440) Roper resonance naturally appears in the nuclear
model with explicit mesons as a structure in the continuum spectrum of
the physical proton which in this calculation is made of a bare nucleon
dressed with a pion cloud.
	\end{abstract}

\section{Introduction}
The N(1440) Roper resonance is a relatively broad nucleon
resonance with the mass of about 1440~MeV and the width
of about 350~MeV~\cite{pdg}. Although its nature is still debated
(see~\cite{HADRON,solution,jose,dynamic} and references therein) one line
of thought is that it consists of a quark core augmented by a meson
cloud. This concurs well with the nuclear model with explicit mesons (MEM)
where the physical nucleon is made of a bare nucleon---the quark
core---dressed with a meson cloud~\cite{sigma,gammapi}.  One
might therefore expect that in MEM the Roper resonance should somehow
reveal itself in the continuum spectrum of the physical proton.

In this contribution we investigate the continuum spectrum of the
physical proton within one-pion MEM in the hope to identify the Roper
resonance and establish the parameters of MEM that are
consistent with the tabulated mass and width of the resonance.
The method we use is calculation of the strength-function of a certain
fictional reaction, where the proton is excitated from the ground state
into continuum, with the subsequent fit of the calculated
strength-function with a Breit-Wigner distribution.

\section{The physical nucleon in MEM}

The MEM is a nuclear interaction model, based on the Schrodinger equation,
where the nucleons do not interact with each other via a potential
but rather emit and absorb mesons~\cite{sigma,gammapi}. The mesons are
treated explicitly on the same footing as the nucleons.

The physical nucleon in MEM is represented by a superposition of
states where the bare nucleon is surrounded by different number of
(virtual) mesons.  In one-pion approximation the physical nucleon is
a superposition of two states: the bare nucleon, and the bare nucleon
surrounded by one pion.  The corresponding wave-function of the physical
nucleon, $\Psi_N$, is a two-component vector,
	\begin{equation}
\Psi_N=
\left(\begin{array}{c}
\psi_0(\vec R) \\
\psi_1(\vec R,\vec r)
\end{array}\right) \;,
	\end{equation}
where $\psi_0$ is the (wave-function of the) state with the bare nucleon
and no pions, $\psi_1$ -- the state with a bare nucleon and one pion,
$\vec R$ is the center-of-mass coordinate of the system, and $\vec r$
is the coordinate between the bare nucleon and the pion.

The Hamiltonian $H$ that acts on this wave-function is a
matrix,
	\begin{equation}
H=\left[\begin{array}{cc}
K_N+m_N & W^\dagger \\
W & K_N+m_N+K_\pi+m_\pi
\end{array}\right] \;,
	\end{equation}
where $K_N, K_\pi$ are kinetic energy operators for the bare nucleon
and the pion, $m_N$ and $m_\pi$ are masses of the bare nucleon and the
pion, and $W$ and $W^\dagger$ are pion emission and absorption operators
(also called the nucleon-pion coupling operators).

The corresponding Schrodinger equation is given as
	\begin{equation}\label{eqn:hpsi}
\left[\begin{array}{cc}
K_N+m_N & W^\dagger \\
W & K_N+m_N+K_\pi+m_\pi
\end{array}\right]
\left(\begin{array}{c}
\psi_0 \\
\psi_1
\end{array}\right)
=E
\left(\begin{array}{c}
\psi_0 \\
\psi_1
\end{array}\right) \;,
	\end{equation}
with the normalization condition
\begin{equation}
\left\langle\Psi|\Psi\right\rangle
=\langle\psi_0|\psi_0\rangle+\langle\psi_1|\psi_1\rangle
=\int_V d^3R|\psi_0|^2
+\int_V d^3R\int_V d^3r|\psi_1|^2=1 \;,
\end{equation}
where $E$ is the energy of the system and $V$ is the normalization volume.

The simplest $W$-operator that is consistent with conservation of isospin,
angular momentum, and parity can be written as
\begin{equation}
W=(\vec\tau\vec\pi)(\vec\sigma\vec r)F(r) \;,
\end{equation}
where $\vec\sigma$ is the vector of Pauli matrices that act on the spin of
the nucleon, $\vec\tau$
is the isovector of Pauli matrices that act on the
isospin of the nucleon\footnote{
The isospin factor $\vec\tau\vec\pi$ is given as
        \begin{equation}
\vec\tau\vec\pi=
\tau_0\pi^0+\sqrt{2}\tau_-\pi^++\sqrt{2}\tau_+\pi^- \;,
        \end{equation}
where $\pi^0$, $\pi^+$, and $\pi^-$ are the physical pions and where
the $\tau$-matrices are given as
       \begin{equation}
\tau_0=\left(
\begin{tabular}{cc}
1 & 0 \\
0 & -1
\end{tabular}
\right)\,,\;
\tau_-=\left(
\begin{tabular}{cc}
0 & 0 \\
1 & 0
\end{tabular}
\right)\,,\;
\tau_+=\left(
\begin{tabular}{cc}
0 & 1 \\
0 & 0
\end{tabular}
\right)\,.
        \end{equation}
},
and where $F(r)$ is a (short-range) form-factor.
The dimension of $W$ is $E/\sqrt{V}$,
therefore it might be of convenience to choose
	\begin{equation}
F(r)=S_wf(r)
	\end{equation}
where $f(r)$ is normalized such that
\begin{equation}\label{eq-f-norm}
\int d^3r r^2f^2(r) = 4\pi\int_0^\infty r^4 f^2(r) dr = 1 \;,
\end{equation}
in which case the strength factor $S_w$ has the dimension of
energy and one can (hopefully) meaningfully compare form-factors of
different shapes\footnote{
The Gaussian form-factor normalized according to~(\ref{eq-f-norm})
is given as
	\begin{equation}\label{eq-gauss-norm}
f(r) =\left(4\pi\frac{3\sqrt{\pi}b_w^5}{2^{11/2}}\right)^{-1/2}
\exp\left(-\frac{r^2}{b_w^2}\right) 
	\end{equation}}.

\section{The semi-relativistic Schrodinger equation for the bare
proton dressed with a pion}
We shall search for the wave-function of the physical proton
in the center-of-mass system in the form
	\begin{equation}\label{eq-psip}
\Psi_{p\uparrow}=
\left(\begin{array}{c}
\frac{p\uparrow}{\sqrt{V}}c_0 \\
(\vec\tau\vec\pi)(\vec\sigma\vec r)\frac{p\uparrow}{\sqrt{V}}
\phi(r)
\end{array}\right)
	\end{equation}
where $p$ is the proton isospin state,
        \begin{equation}
p=\left( \begin{array}{c} 1 \\ 0 \end{array} \right) \;:\;
\tau_0p=p \;,
         \end{equation}
and where $\up$ is the spin-up state,
        \begin{equation}
\up=\left( \begin{array}{c} 1 \\ 0 \end{array} \right) \;:\;
\sigma_0\up=\up \;,
        \end{equation}
and where the dimensionless constant $c_0$ and the function $\phi(r)$
are to be found by solving the Schrodinger equation.
The normalization condition is given as\footnote{
using
	\begin{equation}\label{eq-tau-sigma}
(\vec\tau\vec\pi)^\dagger(\vec\tau\vec\pi)=3 \,,\;
(\vec\sigma\vec r)^\dagger(\vec\sigma\vec r)=r^2 \;.
	\end{equation}
}
	\begin{equation}
\left\langle\Psi_{p\uparrow}|\Psi_{p\uparrow}\right\rangle
=|c_0|^2+3\int d^3r r^2 |\phi(r)|^2
=|c_0|^2+12\pi\int dr r^4 |\phi(r)|^2
= 1 \;.
	\end{equation}

With the ansatz~(\ref{eq-psip}) the Schrodinger equation~(\ref{eqn:hpsi})
turns
into the following system of equations for the constant
$c_0$ and the function $\phi(r)$,
\begin{eqnarray}
\left\{
\begin{array}{l}
m_Nc_0+12\pi\int dr r^4 F(r) \phi(r)
=E c_0 \\
\\
\vec r F(r)c_0
+(K_N+m_N+K_\pi+m_\pi)
\vec r \phi(r)
=E  \vec r \phi(r)
\end{array}
\right. \;.
\end{eqnarray}
It is of advantage to introduce the ``radial'' function $u(r)$,
	\begin{equation}
\phi(r)=\frac{u(r)}{r^2} \;,
	\end{equation}
with the simple boundary condition at the origin,
	\begin{equation}
u(r\to 0)\to0\;,
	\end{equation}
and the normalization condition
	\begin{equation}
|c_0|^2+12\pi\int dr |u(r)|^2 = 1 \;.
	\end{equation}
The Schrodinger equation then becomes
	\begin{eqnarray}
\left\{
\begin{array}{l}
m_Nc_0+12\pi\int dr r^2 F(r) u(r) =E c_0 \\
\\
\vec r F(r)c_0
+(K_N+m_N+K_\pi+m_\pi)
\frac{\vec r}{r^2} u(r)
=E \frac{\vec r}{r^2} u(r)
\end{array}
\right. \;.
	\end{eqnarray}

In the center-of-mass frame the nucleon and the pion have equal
momenta with opposite signs, $-\vec p$ and $\vec p$.
Their (relativistic) kinetic energies are therefore given as
	\begin{equation}
K_N+m_N=\sqrt{m_N^2+\vec p^{\:2}}\;,\,\;
K_\pi+m_\pi=\sqrt{m_\pi^2+\vec p^{\:2}} \;.
	\end{equation}
The momentum as a quantum-mechanical operator
in coordinate space is given as
	\begin{equation}
\vec p=-i\hbar\frac{\partial}{\partial\vec r}\doteq -i\hbar\nabla \;.
	\end{equation}
Correspondingly the kinetic energies operators are
	\begin{equation}
K_N+m_N = \sqrt{m_N^2-\hbar^2\nabla^2}\,,\;
K_\pi+m_\pi = \sqrt{m_\pi^2-\hbar^2\nabla^2} \;.
	\end{equation}
With these kinetic energies the Schrodinger equation turns into
the following system of integro-differential equations,
	\begin{eqnarray}\label{eq-almost}
\left\{
\begin{array}{l}
m_Nc_0+12\pi\int_0^\infty dr r^2 F(r) u(r)
=E c_0 \\
\\
\vec r F(r)c_0 +f_K(\nabla^2)  \frac{\vec r}{r^2}u(r)
=E  \frac{\vec r}{r^2}u(r)
	\end{array}
\right. \;,
\end{eqnarray}
where the function
	\begin{equation}
f_K(x)\doteq \sqrt{m_N-\hbar^2 x}+\sqrt{m_\pi^2-\hbar^2 x}
	\end{equation}
is representable by a Maclaurin series\footnote{
	\begin{equation}
\sqrt{1-x}=1-\frac{x}{2}-\frac{x^2}{8}-\frac{x^3}{16}-\frac{5x^4}{128}
+\dots \;.
	\end{equation}
}.

The single action of the $\nabla^2$ operator on $\frac{\vec r}{r^2}u(r)$
is given as
	\begin{equation}
\nabla^2\left(\frac{\vec r}{r^2}u(r)\right)
=\frac{\vec r}{r^2}\left(u''-\frac{2}{r^2}u\right)
\equiv \frac{\vec r}{r^2} \hat D u
	\end{equation}
where the action of the operator $\hat D$ is defined as
	\begin{equation}
\hat D u \doteq \left(u''-\frac{2}{r^2}u\right) \;.
	\end{equation}
Repeating the action suggests that for any non-negative integer $n$
	\begin{equation}
\left(\nabla^2\right)^n\left(\frac{\vec r}{r^2}u(r)\right)
= \frac{\vec r}{r^2} \hat D^n u \;.
	\end{equation}
Therefore for any function $f(x)$ that can be represented as a power
series the following holds true,
	\begin{equation}
f\left(\nabla^2\right)\left(\frac{\vec r}{r^2}u(r)\right)
= \frac{\vec r}{r^2} f(\hat D) u \;.
	\end{equation}
Inserting this into~(\ref{eq-almost}) gives, finally, the sought
semi-relativistic radial Schrodinger equation for
the physical proton in one-pion MEM,
	\begin{equation}\label{eq-final}
\left\{\begin{array}{l}
m_Nc_0+12\pi\int dr r^2 F(r)u(r) =E c_0 \\
\\
r^2 F(r)c_0 +f_K(\hat D)u(r) = Eu(r)
\end{array}\right. \;,
	\end{equation}
with the boundary condition $u(0)=0$.

\section{Real symmmetric eigenproblem}
The system of equations~(\ref{eq-final}) can be solved numerically
by discretizing the $r$-variable and using the finite-difference
approximations for the derivatives and integrals.

Let us introduce a regular grid,
	\begin{equation}
r_{i=1\dots n}=i\Delta r\,,\; u_i\doteq u(r_i) \,,\; F_i\doteq F(r_i) \;,
	\end{equation}
where $\Delta r$ is the grid spacing.
The integral in the first equation in~(\ref{eq-final}) can then be
approximated as
	\begin{equation}
12\pi\int dr r^2 F(r)u(r) \approx \sum_{i=1}^n \sqrt{12\pi\Delta r}
r_i^2F_i \sqrt{12\pi\Delta r}u_i \;.
	\end{equation}
Introducing the auxiliary tilde-variables,
	\begin{equation}
\tilde F_i \equiv \sqrt{12\pi\Delta r}F(r_i) \,,\; \tilde u_i \equiv
\sqrt{12\pi\Delta r}u(r_i) \;,
	\end{equation}
we turn the system of equations~(\ref{eq-final}) into the following
form,
	\begin{equation}\label{eq-symm}
\left\{\begin{array}{l} m_Nc_0+\sum_{i=1}^n r_i^2\tilde F_i\tilde u_i =
E c_0 \\ \\ r_i^2\tilde F_i c_0 +\sum_{j=1}^n f_K(\hat D)_{ij}\tilde u_j =
E\tilde u_i \end{array}\right. \;,
	\end{equation}
where the numbers $f_K(\hat D)_{ij}$ are the matrix elements of the
matrix representation of the operator $f_K(\hat D)$ on the grid; and
where the ordinary matrix normalization condition applies,
	\begin{equation}\label{eq-matrix-norm}
|c_0|^2+\sum_{i=1}^n |\tilde u_i|^2 = 1 \;.
	\end{equation}

The system of equations~(\ref{eq-symm}) with the normalization
condition~(\ref{eq-matrix-norm}) can be rewritten as a real
symmetric matrix eigenproblem,
	\begin{equation}
\mathcal H \left(\begin{array}{c} c_0 \\ \tilde u_1 \\ \tilde u_2 \\
\vdots \\ \tilde u_n \end{array}\right) =E \left(\begin{array}{c} c_0 \\
\tilde u_1 \\ \tilde u_2 \\ \vdots \\ \tilde u_n \end{array}\right)
	\end{equation}
where the Hamiltonian matrix $\mathcal H$ is given as
	\begin{equation}
\mathcal H=\left[\begin{array}{cccc}
m_N & r_1^2\tilde F_1 & r_2^2\tilde F_2 & \dots \\
r_1^2\tilde F_1 &   &  &  \\
r_2^2\tilde F_2 &  & f_K(\hat D)_{ij} &  \\
\vdots &  &  & 
\end{array}\right] \;.
	\end{equation}

The $\hat D$ operator on the grid can be built using the second order
central finite-difference approximation for the second derivative,
	\begin{equation}
u_i''\approx\frac{u_{i-1}-2u_i+u_{i+1}}{\Delta r^2} \;,
	\end{equation}
which gives
	\begin{equation}
\hat D=\frac{1}{\Delta r^2}
\left[\begin{array}{rrrrr}
-2  & 1 & 0  & 0 &  \dots \\
1   & -2  & 1 & 0 &   \dots \\
0   & 1  & -2 & ~1 &   \dots \\
\vdots & \vdots & \vdots & \vdots & \ddots
\end{array}\right]
+
\left[\begin{array}{rrrrr}
\frac{-2}{r_1^2}  & 0 & 0  & 0 &  \dots \\
0   & \frac{-2}{r_2^2}  & 0 & 0 &   \dots \\
0   & 0  & \frac{-2}{r_3^2} & 0 &   \dots \\
\vdots & \vdots & \vdots & \vdots & \ddots
\end{array}\right] \;.
	\end{equation}
Now the matrix $f_K(\hat D)$ can be built in the following way:
if $\lambda_i$ and $v_i$ are the eigenvalues and eigenvectors of the
$\hat D$ operator,
	\begin{equation}
\hat D v_i = \lambda_i v_i \,,\; i=1\dots n \;,
	\end{equation}
then the matrix representation of any Maclaurin expandable operator
$f(\hat D)$ is given as
	\begin{equation}
f(\hat D)=Vf(\lambda)V^\mathrm{T}
	\end{equation}
where $V$ is the matrix of eigenvectors $v_i$ and where $f(\lambda)$
is a diagonal matrix with diagonal elements $f(\lambda_i)$.

The Hamiltonian matrix built this way has the box boundary conditions
inbuilt,
	\begin{equation}
u(0)=0\,,\; u(R_\mathrm{max})=0 \;,
	\end{equation}
where $R_\mathrm{max}=(n+1)\Delta r$.  Diagonalization of this Hamiltonian
matrix produces one state with the energy below $m_N$ (this state is
the physical proton\footnote{ To be absolutely correct, the mass $m_N$
of the bare nucleon must be larger than the nucleon's physical mass
such that adding the pion binding energy produces the observed physical
mass of the dressed nucleon. However in all our examples later on the
binding energy of the pion is of about 3 to 8\% of the nucleon mass
and therefore in this exploratory investigation we neglect this effect
and assume that $m_N$ equals the physical mass of the nucleon.} --
the bound state of the bare proton and the pion) and
$n$~discretized-continuum states with energies above the pion emission
threshold\footnote{The
pion threshold in these calculations corresponds to a decay into
a bare nucleon and a pion, meaning that in the threshold energy,
$E_\mathrm{th}=m_N+m_\pi$, $m_N$ is strictly speaking the mass of the bare
nucleon that is larger than the physical mass by the binding energy of
the virtual pion. In reality the emitted bare nucleon immediately gets
dressed with another virtual pion which reduces its mass back to the
physical mass, restoring the correct threshold. However this mechanism is
not included in the current one-pion approximation.  However since the
binding of the virtual pion in the present calculations is always small
we assume that $m_N$ is the physical mass.}.  These states decay into
a nucleon and a pion, and the Roper resonance, if any, must be lurking
somewhere there.

\begin{figure}
\caption{The radial wave-functions $\tilde u(r)$:
the ground (bound) state wave-function
$\tilde u_0$, a below-resonance continuum wave-function $\tilde u_3$, and
an on-resonance continuum wave-function $\tilde u_{20}$. The model
parameters are
the ones in the first line in Table~\ref{tab-res},
$\Delta r$=0.08~fm, $n=437$.}
\centerline{\input{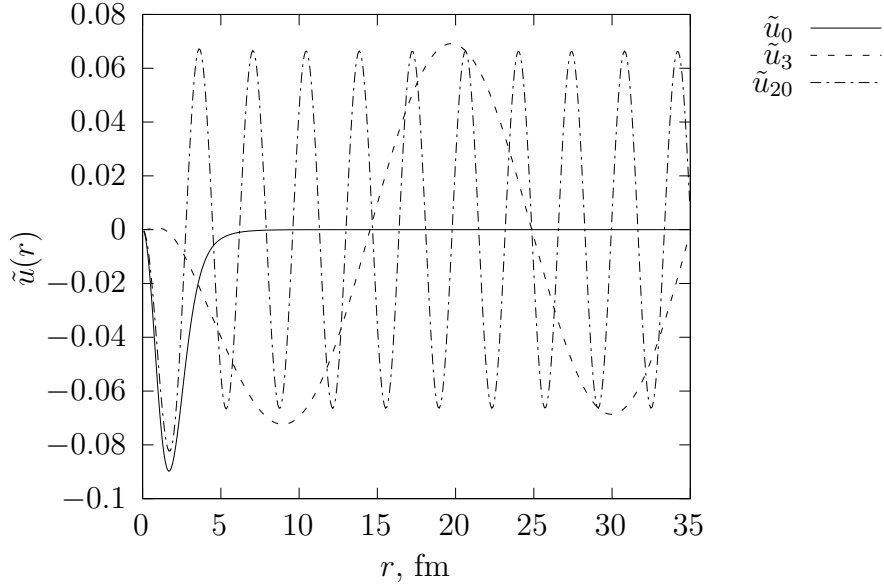}}
\label{fig-u}
\end{figure}

An example of the radial eigenfunctions is shown on Figure~\ref{fig-u}:
the ground state is a localised bound state of the bare nucleon and
the pion; the ``below-resonance'' continuum wave-function shows little
presence at the shorter distances; the ``on-resonance'' wave-function
has a much larger amplitude in the inner region.

\section{The strength-function of a fictional transition}

A resonance is usually identified as a peak in a reaction cross-section
with an approximately Breit-Wigner shape.  The Roper resonance decays
largely (55-75\%~\cite{pdg}) into a nucleon and a pion, therefore it
should be possible to observe the resonance in a reaction where our
dressed proton is excited from the ground state into a continuum spectrum
state which decay exactly into a nucleon and a pion.

Let us consider a reaction where the proton undergoes a transition,
caused by certain operator $\hat X$, from the ground state $\Psi_0$ to
a state in the continuum $\Psi_i$ (which then decays into a nucleon and
a pion).  The amplitude of this quantum transition is given, in the Born
approximation, by the matrix element
	\begin{equation}
M_{i\leftarrow 0} = \langle\Psi_i|\hat X|\Psi_0\rangle \;.
	\end{equation}
Since a resonance should not depend on the way the state $\Psi_i$
is populated, the particular form of the $\hat X$ operator should be
unimportant (as long as the matrix element is not identically zero)
and one can just as well use a fictional operator.  We have chosen the
following matrix element,
	\begin{equation}
M_{i\leftarrow 0} = \int_0^\infty u_i(r)\frac{1}{r^2}u_0(r) dr \;,
	\end{equation}
as experimentation shows that this one produces the best looking
strength-functions (see below).

The cross-section of an excitation reaction into the continuum spectrum
states with energies $E\pm\frac{\Delta E}{2}$
is determined by the so called strength-function, $S(E)$, which is
defined as
	\begin{equation}
S(E) = \frac{1}{\Delta E}\sum_{E_i\in E\pm\frac{\Delta E}{2}}
\left|M_{i\leftarrow 0}\right|^2 \;.
	\end{equation}
In the box-discretized approximation that we use here the strength-function
of a transition into a discretized-continuum state $i$ with the energy $E_i$
is given as
	\begin{equation}\label{eq-S}
S(E_i) = \frac{1}{\Delta E_i}\left|M_{i\leftarrow 0}\right|^2
	\end{equation}
where $\Delta E_i=E_{i+1}-E_i$.

One can determine the parameters of a resonance, the mass and the width,
by applying a Breit-Wigner fit to the strength-function~\cite{otrap}.
Unfortunately the Roper resonance is broad and is located close to
threshold making it necessary to use a width that is energy-dependent.
We shall use the following phenomenological parametrization of the
strength-function (cf.~\cite{sill}),
	\begin{equation}\label{eq-BW}
S(E) \propto
\frac{(M\tilde\Gamma)^2}{(E^2-M^2)^2+(M\tilde\Gamma)^2}
\theta(E-E_\mathrm{th}) \;,
	\end{equation}
where
	\begin{equation}
\tilde\Gamma = \Gamma\left(
\frac{E^2-E_\mathrm{th}^2}{M^2-E_\mathrm{th}^2}\right)^p \;,
	\end{equation}
$\theta$ is the step function, $E_\mathrm{th}$ is the threshold energy,
and where the mass $M$, width $\Gamma$, and power $p$ are the fitting
parameters ($p\sim 1$).

\section{Results}
We use the Gaussian form-factor in the normalized
form~(\ref{eq-gauss-norm}),
	\begin{equation}\label{gauss-F}
F(r) = S_w\left(4\pi\frac{3\sqrt{\pi}b_w^5}{2^{11/2}}\right)^{-1/2}
\exp\left(-\frac{r^2}{b_w^2}\right) \;,
	\end{equation}
where the strength, $S_w$, and the range, $b_w$, are the model parameters
that are varied to reproduce the given mass $M$ and width $\Gamma$ of the
resonance.

\begin{figure}
\caption{The calculated strength-functions~(\ref{eq-S}) together with their
Breit-Wigner fits~(\ref{eq-BW}) for the parameter sets from
Table~\ref{tab-res}.}
\label{fig-sfun}
\centerline{\input{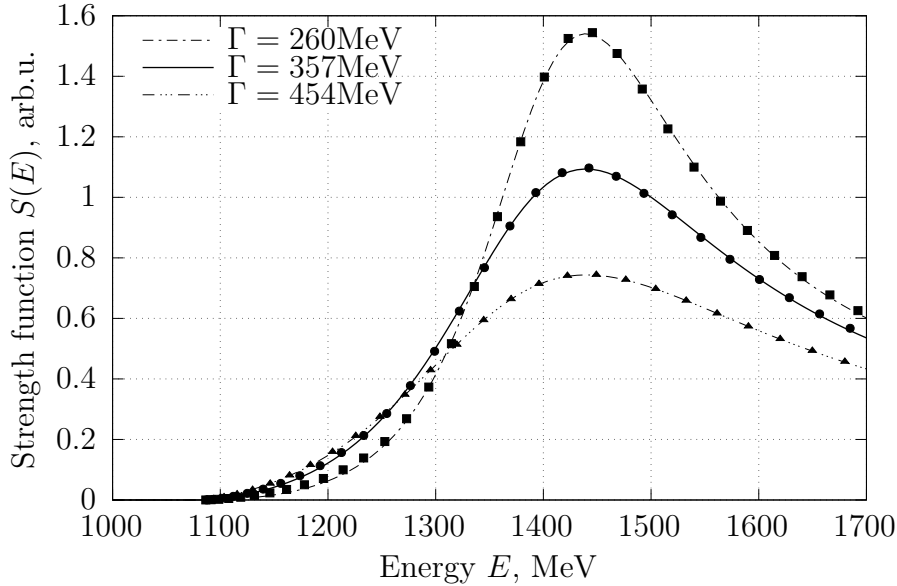}}
\end{figure}

Exploratory calculations show that it is possible to reproduce the
tabulated mass of the Roper resonance, 1440~Mev, and any given width
within the tabulated limits, 250-450~MeV, with a relatively narrow
range of the model parameters as indicated in Table~\ref{tab-res} and
on Figure~\ref{fig-sfun}.

\begin{table}
\caption{The results of the exploratory calculations: $S_w$ and $b_w$ are
the strength and the range parameters of the Gaussian
form-factor~(\ref{gauss-F}), $M$ and
$\Gamma$ are the mass and the width of the resonance from the Breit-Wigner
fit~(\ref{eq-BW}) to the strength-function~(\ref{eq-S}), $B$ is the
binding energy of the virtual pion,
and $N_\pi=\langle\psi_1|\psi_1\rangle=(1-c_0^2)$ is the contribution
of the state with the virtual
pion to the total norm of the physical proton wave-function.
~\\}
\centerline{\begin{tabular}{|c|c|c|c|c|c|}
\hline
$S_w$, MeV & $b_w$, fm & $M$, MeV & $\Gamma$, MeV & $B$, MeV & $N_\pi$, \% \\
\hline
  99       &  1.50     & 1439     &  260          &   72     &  16  \\
  80       &  1.42     & 1440     &  357          &   48     &  11  \\
  63       &  1.35     & 1440     &  454          &   30     &  8  \\
\hline
\end{tabular}}
\label{tab-res}
\end{table}

\section{Conclusion}
We have shown that in the nuclear model with explicit mesons (MEM) the
Roper resonance appears as a structure in the continuum spectrum of the
physical proton (in one-pion MEM the physical proton is made of
a bare nucleon dressed with a pion). We have established the range of the
model parameters (the strength and the range of the nucleon-pion
coupling operator) that reproduce the tabulated mass and width of the
resonance.  With these parameters the pion component in the physical
proton takes about 10\% of the norm of the wave-function and reduces
the mass of the bare proton by about 5\%.  These numbers are much
smaller then the corresponding estimates from the pion photo-production
cross-section calculation~\cite{gammapi}.

\end{document}